\documentclass{kapproc}

\normallatexbib

\begin{document}
\input epsf

\articletitle{Sketching the inflaton potential}

\author{C\'esar A. Terrero-Escalante}
\affil{Instituto~de~F\'{\i}sica,~UNAM,%
~Apdo.~Postal~20-364,~01000,~M\'exico~D.F.,~M\'exico}
\email{cterrero@fis.cinvestav.mx}

\chaptitlerunninghead{Sketching the inflaton potential}

\begin{abstract}

Based on solutions of the Stewart-Lyth inverse problem
it is argued that in the CMB data analysis
the parametrization of the
primordial spectra from inflation must include the `running'
of both, scalar and tensorial, spectral indices if information
beyond the exponential potential model is wanted to be detected.

\end{abstract}


\section*{Introduction}

Usually, to understand a grownup behavior it is necessary to look back into
his childhood.
This seems to be also the case in Cosmology.
(See Refs.~\cite{inflation} for reviews and references
on the topics mentioned in this introduction).

The eldest picture we have of our universe comes from
the times when it was about $10^5$ years old. In that epoch matter became
cold enough for radiation to decouple and almost freely travel
across the space up to the present. Since cosmic time ticks logarithmically
and the current age of the universe is estimated to be about
$10^{10}$ years,
this nearly uniform cosmic background of radiation is a tidy picture
of the young universe.

Nuclear physics allow us to trace the cosmic history
back to the times when the lightest chemical elements were synthesized. The
predicted abundances of these elements can be matched with current surveys,
providing
evidence about
the universe when it was a three seconds old kid.

High-energies physics as described by the
Standard Model of Particles
allow us
to glance a little bit further into the past, but the description of the
events happening immediately after the birth of our universe still remains
highly speculative.
The main problem here is the lack of a consistent and tested
theory,
 and the practical impossibility of reproducing the relevant events in
laboratory conditions because of the very high energies involved.
It is, therefore, quite stimulating that the inflationary paradigm links the
physics in the very early universe to the current cosmological state.

A period of rapid accelerated expansion of the universe just after its birth
can explain several features of the currently observed universe like its age,
its size and its topology. Perhaps the more attractive
feature that
can be explained in this framework are the initial conditions for our own
existence. In a perfectly homogeneous universe there are not chances for
such complex structures like we are to arise.
At the epoch when non-relativistic matter
dominated over the relativistic one (or radiation),
inhomogeneities acting like
perturbations in the gravitational
potential were required to seed the
formation of galaxies through gravitational instability.
Ultimately, this process
led to the formation of solar systems and
planets like the Earth. Those perturbations produced
when the pressure of the radiation was still high enough
to compensate gravity attraction
induced
an oscillatory motion of expansions and contractions. This motion led to
inhomogeneities in the background temperature which we observe
today as anisotropies
in the cosmic microwave background (CMB) radiation. Therefore, to describe
the formation of galaxies and the CMB anisotropies spectrum it is necessary
to set the spectrum of the corresponding
 initial perturbations of the gravitational
potential. Given the very special characteristics of this
primordial spectrum,
the inflationary scenario is the most widely accepted mechanism for setting
these initial conditions \cite{inflation,CMBdata}.
In an inflationary universe quantum
fluctuations of matter and space-time are stretched by the expansion up to
cosmological scales well beyond the distance within which causal interaction
can take place. After the end of inflation these fluctuations reenter the
causal horizon producing perturbations in the gravitational potential.

It is remarkable that the kind of primordial perturbation spectra generated
in the simplest version of the inflationary scenario, the single scalar field
scenario, fits quite well as the seeds for the CMB anisotropies
\cite{inflation,CMBdata}.
In this scenario the expansion is driven by the potential
energy of the, so called, inflaton field. This way, if it is assumed, as we do
here, that the inflaton physics closely correspond to the actual scenario
of the very early universe, then recalling the earliest memories of our
universe is equivalent to sketching the inflationary potential.

The aim of this contribution is to discuss how the Stewart-Lyth inverse
problem \cite{Ayon-Beato:2000xx}
can be used as a powerful tool for drawing a photo-robot of the
inflaton potential.

\section{Stewart-Lyth inverse problem}
\label{sec:slip}

To describe the inflationary dynamics and the corresponding perturbations it
has proved to be useful to introduce the set of horizon flow functions
\cite{Schwarz:2001vv}:
\begin{equation}
\epsilon_0 \equiv \frac{{d_{\rm H}}(N)}{d_{\rm Hi}},
\end{equation}
and,
\begin{equation}
\label{eq:Hjf}
\epsilon_{m+1} \equiv {{\rm d} \ln |\epsilon_m|
\over {\rm d} N}, \qquad m\geq 0.
\end{equation}
where
$d_{\rm H} \equiv 1/H \equiv a/\dot{a}$ denotes the Hubble distance,
with $d_{\rm Hi}$ evaluated
at some initial time $t_i$
and dot stands for differentiation with respect to cosmic time.
The scale factor $a$ measures the expansion
of the spatial volume, and
$N\equiv \ln(a/a_i)$ is the efolds number.
The first
horizon flow function $\epsilon_1$ can be written in several useful ways:
\begin{equation}
\label{eq:e1Gen}
\epsilon_1 \equiv {{\rm d} \ln d_{\rm H}\over {\rm d} N} =
\dot{d_{\rm H}} =
\frac32\frac{\rho+p}\rho =
3\,\frac{T}{T + V} =
\kappa \frac{T}{H^2} \, ,
\end{equation}
where
$\rho$ and $p$ are, respectively, the energy density and the pressure
in a universe dominated by the potential energy, $V(\phi)$, of the
inflaton field, $\phi$. $T\equiv\dot{\phi}^2/2$,
$\kappa = 8\pi/m_{\rm Pl}^2$ is the
 Einstein constant and $m_{\rm Pl}$ is the Planck mass.
Inflation happens for $0\leq\epsilon_1 < 1$.
For $m>1$, $\epsilon_m$ may take
any real value.

At the high energies corresponding to the inflationary period, the
inflaton energy density
and the space-time metrics
undergo quantum fluctuations. These are the fluctuations
that stretched by the expansion reentered the Hubble horizon, $d_H$,
close to the point of equal density of matter and radiation,
to seed
the formation of large scale structure
and the CMB anisotropies.
The spectra of these seeds can be
parametrized by the series:
\begin{eqnarray}
\label{eq:SExp}
\ln A_S^2(k)&=&\ln A_S^2(k_*) + \Delta(k_*)\ln\frac{k}{k_*}
+ \frac12\frac{d\Delta(k)}{d\ln k}\vert_{k=k_*}\ln^2\frac{k}{k_*} + \cdots
\,,\\
\label{eq:TExp}
\ln A_T^2(k)&=&\ln A_T^2(k_*) + \delta(k_*)\ln\frac{k}{k_*}
+ \frac12\frac{d\delta(k)}{d\ln k}\vert_{k=k_*}\ln^2\frac{k}{k_*} + \cdots
\,,
\end{eqnarray}
where $A_S$ and $A_T$ stand respectively for the normalized amplitudes of the
density (scalar) and
metrics (tensor) perturbations, and $k_*$ is the wavenumber corresponding to
a pivotal length scale.
Functions $\Delta(k)$ and $\delta(k)$ will be called here
the scalar and tensorial spectral indices respectively. Their derivatives
with respect to $\ln k$ are known as the `running' of the spectral
indices.
In the procedure of fitting the CMB data
the order where series (\ref{eq:SExp}) and (\ref{eq:TExp}) are truncated
is determined by the precision of the observations.
It is worthy to mention that a careful analysis of the perturbations
generated during inflation shows that the spectra of such perturbations
while reentering the Hubble horizon are generically almost scale-invariant
(see Ref.~\cite{inflation} for details). With this regard, one can
assume that each higher term in series (\ref{eq:SExp}) and (\ref{eq:TExp})
is smaller than the corresponding lower order term.

Typically, most
of CMB data analyses have neglected
the possible effects of the
primordial gravitational wave spectrum \cite{CMBdata}.
The tensor contribution to the CMB spectrum
can be parametrized in terms of the quantity
\begin{equation}
\label{eq:r}
r \equiv \alpha\frac{A_T^2}{A_S^2}\, ,
\end{equation}
representing the relative amplitudes
of the tensor and scalar perturbations,
where the constant, $\alpha$, depends on the
particular normalization of the spectral amplitudes that is chosen.
In the last few years, however,
there has been a growing recognition that
the role of the tensor perturbations
deserves more attention when determining the
best--fit values of the
cosmological parameters
(For a recent review, see, e.g., Ref. \cite{CMBdata}).

The Stewart-Lyth inverse problem (SLIP) was
introduced in Ref.~\cite{Ayon-Beato:2000xx}
as a method for finding the
inflaton potential using information on the functional form of the spectral
indices. With this aim, the following non-linear differential equations
must be solved:
\begin{eqnarray}
2C\epsilon_1 \hat{\hat{\epsilon_1}}-(2C+3)\epsilon_1 \hat{\epsilon_1}
-\hat{\epsilon_1}
+\epsilon_1 ^{2}+\epsilon_1 +\Delta &=&0\,, \label{eq:MSch1} \\
2(C+1)\epsilon_1 \hat{\epsilon_1}-\epsilon_1 ^{2}-\epsilon_1 -\delta &=&0\,,
\label{eq:MSch2}
\end{eqnarray}
where $C=-0.7293$ and
a circumflex accent denotes differentiation with respect to
$\tau\equiv \ln H^2$. These equations are derived from next-to-leading
order expressions for the spectral indices in terms of the horizon flow
functions (\ref{eq:Hjf}) \cite{Ayon-Beato:2000xx}.
In Ref.~\cite{Terrero-Escalante:2001ni}
it was shown that to this order $\delta\leq0$.
The validity of the conclusions to be
drawn here are constrained by the next-to-leading order precision.

Then, having an expression for $\tau(\epsilon_1)$ (which is typically the
form of the solutions to Eqs.~(\ref{eq:MSch1}) and (\ref{eq:MSch2})) the
corresponding
inflaton potential is given by the parametric function
\cite{Terrero-Escalante:2001ni},
\begin{equation}
V(\phi)= \left\{
\begin{array}{c}
\phi(\epsilon_1)\, ,  \\
V(\epsilon_1)\, ,
\end{array}
\right.
\label{eq:FVphie}
\end{equation}
where,
\begin{equation}
V(\epsilon_1)= \frac{1}{\kappa}\left(3-\epsilon_1\right)
\exp\left[\tau(\epsilon_1)\right]\,,
\label{eq:PotentialT}
\end{equation}
and,
\begin{equation}
\phi(\epsilon_1)= -\frac{2(C+1)}{\sqrt{2\kappa}}
\int\frac{\sqrt{\epsilon_1}d\epsilon_1}{\epsilon_1^2+\epsilon_1+\delta}
+ \phi_0\, .
\label{eq:Phie}
\end{equation}
Here $V_0$ and $\phi_0$ are integration constants.
The SLIP solutions are constrained by the following conditions
\cite{Terrero-Escalante:2001ni}:
\begin{equation}
\left\{
\begin{array}{rcl}
\hat{\epsilon_1}\frac{d\phi}{d\epsilon_1}&<&0\, ,  \\
\hat{\epsilon_1}\frac{dV}{d\epsilon_1}&>&0\, .
\end{array}
\right.
\label{eq:epsConds}
\end{equation}

In order to link the SLIP solution with the primordial spectra given by
the series (\ref{eq:SExp}) and (\ref{eq:TExp})
it proved to be useful to rewrite
Eqs.~(\ref{eq:MSch1})
and (\ref{eq:MSch2}) by converting from derivatives with respect to $\tau$ to
derivatives with respect to $\ln k$, with $k$
corresponding to the wavelength crossing the Hubble
horizon by the first time, i.~e., $k=aH$ \cite{Ayon-Beato:2000ga}.

\section{Constraining the inflaton potential}

The Stewart-Lyth inverse problem
has two
strong drawbacks.
First, the full power of
this procedure can be used only when information on the functional forms
of both spectral indices is available. Unfortunately, this information
is rather difficult to be directly obtained from observations.
In addition,
simple functional forms of the spectral indices involve great
difficulties while solving the SLIP.
In spite of these drawbacks, there is an
alternative way of using the SLIP related expressions.
This method allows
to test for the internal consistency in the procedure of fitting the CMB data
by finding and describing the inflaton potential corresponding to the
assumptions on the primordial perturbations used in that procedure.
Thus,
the SLIP can help in constraining the possible inflaton potentials by
linking distinct features of the power spectra to the inflationary dynamics.

Historically, the first
analytical calculations
of the form of the initial conditions required for galaxy formation
due to Harrison and Zel'dovich (see Ref.~\cite{inflation} for details)
yield
scale-invariant spectra, i.e.,
constant amplitudes ($\Delta=\delta=0$).
Current precision of the CMB measurements still allows for this kind
of spectra to be used as initial conditions \cite{CMBdata}. However,
in the inflationary scenario some degree of scale dependence is
necessarily present if a dynamical
mechanism for ending the accelerated expansion
acted in the very early universe
in order to recover the success of the Hot Big Bang Theory
as a sequence of a radiation and a matter dominated universes.

Since the contribution of the tensor modes to the CMB
anisotropies is typically very small, one can wonder whether the actual
inflaton potential could produce perturbations spectra where the
scale dependence is entirely contained in the scalar component, i.~e.,
$\delta=0$ and $\Delta=\Delta(k)$.

The corresponding SLIP was solved in Ref.~\cite{Terrero-Escalante:2001ni}.
The inflaton potential has the form,
\begin{equation}
\label{eq:Vdelta0}
V(\phi)=V_0\frac{3-\tan^2\left[\frac{\sqrt{2\kappa}}
{4(C+1)}(\phi-\phi_0)\right]}
{\cos^{4(C+1)}\left[\frac{\sqrt{2\kappa}}
{4(C+1)}(\phi-\phi_0)\right]}\, .
\end{equation}
Nevertheless, the analysis of the behavior of the scalar index,
\begin{eqnarray}
\Delta(k)\!\!\!\!\!&=&\!\!\!\!\!
\frac1{8\left(C+1\right)^2}\left\{
\left(\frac k{k_0}\right)^{1/\left(C+1\right)}
-\sqrt{\left( \frac k{k_0}\right) ^{1/\left(
C+1\right) }\left[ \left( \frac k{k_0}\right) ^{1/\left( C+1\right)
}-4\right] }\right\}  \nonumber \\
&\times&\!\!\!\!\!
\left\{ \left( \frac k{k_0}\right) ^{1/\left( C+1\right) }
+2C-\sqrt{\left(\frac k{k_0}\right)^{1/\left(C+1\right)}
\left[\left(\frac k{k_0}\right)^{1/\left(C+1\right)}-4\right] }\right\}
,
\label{eq:Dk0}
\end{eqnarray}
corresponding to the potential (\ref{eq:Vdelta0}),
shows that, for large $k$, Eq.~(\ref{eq:Dk0}) converges to
$\Delta=1/2(C+1)\approx 1.85$,
which is too far from values allowed by theory
and observations \cite{inflation,CMBdata}.

Therefore, a correct parametrization of the inflationary spectra
must take into account terms beyond the constant ones in both series
(\ref{eq:SExp}) and (\ref{eq:TExp}). The next step is, then, to
consider spectra slightly tilted from scale invariance, i.e.,
with constant and close to zero spectral indices. In that case the spectra
take on a power-law form. It is well known that an exponential potential
produces exactly such perturbation spectra. This scenario is
called power-law inflation because $a\propto t^p$, where $p\gg 1$ is a
constant \cite{PLinfl}.
Using Eqs.~(\ref{eq:e1Gen}),
it is easy to check that for this model, $\epsilon_1=1/p$.

We proved in Ref.~\cite{Terrero-Escalante:2001ni} that,
even to next-to-leading order,
the only SLIP solution corresponding to these conditions on both spectra,
tensorial and scalar, is power-law inflation. If, in the same mood as
it was done while analyzing the case $\delta=0$, the condition of a power-law
spectrum is relaxed by allowing scale dependence only for the scalar index,
then it will be obtained that, though power-law inflation is still a
trivial solution, other SLIP solutions arise \cite{Terrero-Escalante:2001ni}.
(Hereafter the involved formulas for
the SLIP solutions are omitted.
For details see the cited references).
These solutions
converge to power-law inflation; the scalar index converges to a
constant realistic value from above or from below, depending on the initial
conditions. Since the depart from power-law behavior takes place at large
angular scales where the cosmic variance is dominant, then it will be
very difficult from the observational point of view to distinguish
between these SLIP solutions and power-law inflation.

This way, if one wants to move beyond the power-law scenario, it is
necessary to include the running of both spectral indices in the
parametri\-zation given by Eqs.~(\ref{eq:SExp}) and (\ref{eq:TExp}).
It is reasonable, for instance, to consider that the scale dependence
of the tensorial spectral index is distinctly perceived only up to
next-to-leading order in terms of $\epsilon_1$, consistently with
the approximation used to derive Eqs.~(\ref{eq:MSch1}) and (\ref{eq:MSch2}).
We solved the SLIP with the ansatz \cite{Terrero-Escalante:2001rt},
\begin{equation}
\label{eq:dansatz}
\delta(\epsilon_1)=-\left[(1+a)\epsilon_1^2+(1+b)\epsilon_1+c\right]
\, ,
\end{equation}
where $a,b,c$ are real
numbers. It was found that for $b^2>4ac$, power-law inflation is again
an attractor of the corresponding inflationary dynamics.
For $b^2<4ac$, power-law inflation is no longer an attractor
but a transient regime of the dynamics.
Obviously, the spectra depart from a power law before and after
the quasi power-law behavior. Those perturbations produced before
the quasi power-law regime are imprinted in the very large angular
scales, being the detection of their signatures difficult because of the
cosmic variance. On the other hand,
since inflation ends
very fast after the quasi power-law regime is left behind,
the scales crossing the
horizon at that time
were extraordinarily small and reentered
it back immediately with not relevant effect. If the
actual inflaton potential is close to this model, it will be also
very difficult to observe any difference from the exponential potential.
Then, even including the running of both spectral indices in the
parametrization of the primordial spectra could be not sufficient to
get ride of the power-law bias.

There are cases where power-law inflation is a repellor
of the inflationary dynamics.
In
Ref.~\cite{Terrero-Escalante:2001du} the SLIP was solved with the
condition of constant tensor to scalar ratio, $r$. This was partially
motivated by the possibility, in the near future, of estimating a
constant central value for $r$ from the observation of the
CMB polarization \cite{inflation,CMBdata}.
From Eqs.~(\ref{eq:SExp}), (\ref{eq:TExp}),
(\ref{eq:MSch1}) and (\ref{eq:MSch2}),
it is simple to show that the condition
$r=constant$ is equivalent to $\Delta(k)=\delta(k)$. Using this, after
adding Eqs.~(\ref{eq:MSch1}) and (\ref{eq:MSch2}),
\begin{equation}
\label{eq:Meq}
2C\epsilon_1\hat{\hat{\epsilon_1}}-(\epsilon_1+1)
\hat{\epsilon_1} = 0\, .
\end{equation}
Obviously, a trivial solution for this case is
$\epsilon_1=1/p=constant$,
corresponding to power-law inflation. The remaining
SLIP solutions depart very quickly from this power-law solution.
In one case $\delta$ grows and becomes positive,
thus indicating a breakdown in the next--to--leading order
analysis. For the other case, $\delta$ begins to evolve
extremely rapidly, probably indicating that
the running of the spectral index, $d \Delta/ d\ln k$,
becomes too large. Either way, observational
constraints are difficult to satisfy.
Thus, the potential in the region open to observation
must be sufficiently close to the exponential (power-law
inflation) model.

\section{Conclusions}

The quality of the photo-robot of the inflaton potential depends crucially
on the order of the series used to parametrize the primordial spectra
while fitting the CMB anisotropies spectrum. In turn, that order depends on
the precision of the available observational data.

If the inflaton potential is suspected to differ from the exponential
potential corresponding to power-law inflation, then quadratic terms of
both scalar and tensor modes are necessary (but perhaps not sufficient)
to observe the differences.

If, while fitting the CMB data, the running of both spectral indices
happen to be distinctly different from zero, then the constant value
of the tensor to scalar ratio as determined from the CMB polarization
likely will not be characteristic of a wide range of scales.

When this contribution was ready to be submitted, a paper
by Leach et al. \cite{Leach:2002ar} was posted at Los Alamos including a very
interesting discussion on the parametrization of the primordial
spectra from inflation.

\begin{acknowledgments}

The author wish to thank the
organizers of the Meeting by their invitation to participate.
The author was partially supported by
the CONACyT grant 38495--E and the
Sistema Nacional de Investigadores (SNI).
\end{acknowledgments}


\begin{chapthebibliography}{1}

\bibitem{inflation}
A. D. Linde,
{\em Particle Physics and Inflationary Cosmology},
Chur: Harwood, 1990, pp. 362.;
A. R. Liddle and D. H. Lyth,
{\em Cosmological Inflation and Large-Scale Structure},
Cambridge: Univ. Press, 2000, pp. 400.

\bibitem{CMBdata}
 W.~Hu and S.~Dodelson, {\tt arXiv:astro-ph/0110414} (2001).

\bibitem{Ayon-Beato:2000xx}
E.~Ayon-Beato, A.~Garcia, R.~Mansilla and C.~A.~Terrero-Escalante,
{\em Phys.\ Rev.}\  {\bf D62}, (2000) 103513.
[arXiv:astro-ph/0007477].

\bibitem{Schwarz:2001vv}
D.~J.~Schwarz, C.~A.~Terrero-Escalante and A.~A.~Garcia,
{\em Phys.\ Lett.}\ {\bf B517}, (2001) 243.
[arXiv:astro-ph/0106020].

\bibitem{Terrero-Escalante:2001ni}
C.~A.~Terrero-Escalante, E.~Ayon-Beato and A.~A.~Garcia,
{\em Phys.\ Rev.}\ {\bf D64}, (2001) 023503.
[arXiv:astro-ph/0101522].

\bibitem{Ayon-Beato:2000ga}
E.~Ayon-Beato, A.~Garcia, R.~Mansilla and C.~A.~Terrero-Escalante,
{\tt arXiv:astro-ph/0009358} (2000).

\bibitem{PLinfl}  F.\ Lucchin and S.\ Matarrese,
{\em Phys.\ Rev.}\ {\bf D32}, (1985) 1316.

\bibitem{Terrero-Escalante:2001rt}
C.~A.~Terrero-Escalante and A.~A.~Garcia,
{\em Phys.\ Rev.}\ {\bf D65}, (2002) 023515.
[arXiv:astro-ph/0108188].

\bibitem{Terrero-Escalante:2001du}
C.~A.~Terrero-Escalante, J.~E.~Lidsey and A.~A.~Garcia,
{\tt arXiv:astro-ph/0111128} (2001).

\bibitem{Leach:2002ar}
S.~M.~Leach, A.~R.~Liddle, J.~Martin and D.~J.~Schwarz,
{\tt arXiv:astro-ph/0202094} (2002).

\end{chapthebibliography}
\end{document}